# Time Series, Stochastic Processes and Completeness of Quantum Theory

Marian Kupczynski

*Department of Mathematics and Statistics, University of Ottawa, 585,av. King-Edward,Ottawa.Ont.K1N 6N5*
*and*
*Département de l'Informatique, UQO, Case postale 1250 ,succursale Hull, Gatineau. Quebec, Canada J8X 3X 7*

**Abstract.** Most of physical experiments are usually described as repeated measurements of some random variables. Experimental data registered by on-line computers form time series of outcomes. The frequencies of different outcomes are compared with the probabilities provided by the algorithms of quantum theory (QT). In spite of statistical predictions of QT a claim was made that it provided the most complete description of the data and of the underlying physical phenomena. This claim could be easily rejected if some fine structures, averaged out in the standard descriptive statistical analysis, were found in time series of experimental data. To search for these structures one has to use more subtle statistical tools which were developed to study time series produced by various stochastic processes. In this talk we review some of these tools. As an example we show how the standard descriptive statistical analysis of the data is unable to reveal a fine structure in a simulated sample of AR (2) stochastic process. We emphasize once again that the violation of Bell inequalities gives no information on the completeness or the non locality of QT. The appropriate way to test the completeness of quantum theory is to search for fine structures in time series of the experimental data by means of the purity tests or by studying the autocorrelation and partial autocorrelation functions.

## INTRODUCTION

The algorithms provided by quantum theory (QT) allow finding the probability distributions of experimental data. In some experiments beams of "identical physical systems" are prepared and their interaction with a measuring apparatus produces various time series of outcomes. In other experiments a "single physical system" placed in a trap is probed by a laser beam and the measurements are made. The state of the system in the trap is reset to the initial state and the procedure is repeated. Any single experimental outcome is not predictable only the statistical regularities are observed and compared with the predictions of QT.

It is well known since the explanation of Bertrand's paradox [1-3] that a probability distribution is neither a property of a coin nor a property of a flipping device. It is only the characteristic of the whole random experiment: "flipping this particular coin with that particular flipping device". Therefore the claim that QT provides the complete description of individual physical systems can not be correct and by no means the quantum state vector can be treated as an attribute of an individual physical system.

For Einstein and Blokhintsev QT was only a statistical theory. Bohr wisely insisted on the wholeness of physical experiments underlying the epistemological, algorithmic and contextual character of the theory.

A contemporary statistical interpretation of QT is contextual as it should be. The arguments in favor of this interpretation and many other references may be found in Balletine [4, 5], Accardi [6-8], Khrennikov [9-11] and in [3, 12-16].

Statistical description provided by QT leaves open a question whether it is possible to find a deterministic sub-quantum description of the phenomena in which the uncertainty of the individual outcomes would result only from the lack of control of some "hidden parameters" describing the "physical systems" and the measuring devices.

Even if personally I am not an advocate of hidden variable models because of their ad hoc character and restricted applicability I was impressed by the event based corpuscular model [17] presented during this conference by Hans de Raedt. This model gives a unified description of all interesting optical experiments. It is a nice example of a contextual model which event after event builds the statistical distribution of counts consistent with the predictions of QT without using the Maxwell theory or the quantum mechanics. It gives an additional justification for the search of fine structures in the physical data as well as for the search of new alternative more detailed and non standard models of physical phenomena.

Many years ago we observed that if the hidden variables existed then all pure quantum ensembles would be mixed statistical ensembles with respect to these variables [18]. Since mixed statistical ensembles differ from pure statistical ensembles the difference could be discovered with help of the purity tests [19, 20].

In couple of recent papers we went a step further and we noticed that even the predictable completeness of QT has not been tested carefully enough [3,14,21]. Namely we pointed out that any hypothetical fine structure in the time series of the experimental data, if it existed, it would be averaged out when empirical histograms were constructed from the data and compared with the probability distributions provided by QT. If some reproducible unexpected fine structures were discovered in the data it would be a significant discovery and a decisive proof of incompleteness of QT. In this paper which is a continuation of the paper [21] we review some statistical tools which are used to detect autoregressive structures in a time series of data [22].

## TIME SERIES THEORY

A time series it is a family of random variables $\{Z_t\}$ where $t=0,1..$ for simplicity. Time series is stationary if $E(Z_t)=\mu$, $var(Z_t)=\sigma^2$ and auto covariance function $\gamma(k)$ at lag k does not depend on t where $\gamma(k) = cov(Z_t, Z_{t+k}) = E(Z_t -\mu, Z_{t+k} -\mu)$.

A white noise it is a time series $\{a_t\}$ where $a_t$ are normal independent and identically distributed (i.i.d) random variables with zero mean and variance $\sigma^2$.

The autocorrelation function $\rho(k)$ at lag k is defined as $\rho(k)= \gamma(k) / \gamma(0)$ and it is easy to see that $\rho(0)=1$, $\rho(k) =\rho(-k)$ and $|\rho(k)|\leq 1$  $k=0,\pm 1, ..$

It is useful to introduce the following operators: $BZ_t = Z_{t-1}$ and $\nabla = I-B$.

The important models of time series, which were studied extensively [22], are so called autoregressive integrated moving average models ARIMA (p,d,q):
$\Phi(B)(I-B)^d Z_t = \Theta_0 + \Theta(B)a_t$ where $\Phi(B)=I-\Phi_1 B- \Phi_2 B^2-\ldots-\Phi_p B^p$ and
$\Theta(B)=I- \Theta_1 B- \Theta_2 B^2- \ldots- \Theta q B^q$.

In this paper we concentrate on simple autoregressive models
AR(p)= ARIMA(p,0,0) with µ=0 given by the equation:

$$Z_t - \Phi_1 Z_{t-1} - \Phi_2 Z_{t-2} - \ldots - \Phi_p Z_{t-p} = a_t \tag{1}$$

A general solution of this difference equation is a sum of homogeneous and particular solutions: $Z_t = Z_h(t) + Z_p(t)$. Guessing $Z_h(t) = G^t$ we find that allowed values $G_i$ of G can be found as the roots of the following polynomial equation: $\Phi(G^{-1})=0$

If $|G_i|<1$ then AR (p) is stationary and if the polynomial equation has p different complex roots $G_i$ we find:

$$Z_h(t) = C_1 G_1^t + C_2 G_2^t + \ldots + C_p G_p^t \tag{2}$$

Thus $Z_h(t)$ decays to zero as a sum of exponentials and/or damped sine functions and $Z_t \approx Z_p(t)$ if *t* grows.

Similar behavior has the autocorrelation function ρ(k) which is usually denoted in statistical packages as ACF. Since $E(Z_{t+k},a_t)=0$ we get for k>0 the following homogeneous difference equation similar to Eq. 1:

$$\rho(k) - \Phi_1 \rho(k-1) - \Phi_2 \rho(k-2) - \ldots - \Phi_p \rho(k-p) = 0 \tag{3}$$

Thus the general solution of this equation is given by:

$$\rho(k) = C_1 G_1^k + C_2 G_2^k + \ldots + C_p G_p^k \tag{4}$$

and ρ(k)=ACF is a quickly decaying function when k increases.

From Eq, 3 one obtains so called Yule-Walker matrix equations:

$$\begin{bmatrix} 1 & \rho_1 & \rho_2 \ldots & \rho_{p-1} \\ \rho_1 & 1 & \rho_1 \ldots & \rho_{p-2} \\ \ldots & \ldots & \ldots & \ldots \\ \rho_{p-1} & \rho_{p-2} & \rho_{p-3} \ldots & 1 \end{bmatrix} \begin{bmatrix} \Phi_1 \\ \Phi_2 \\ \ldots \\ \Phi_p \end{bmatrix} = \begin{bmatrix} \rho_1 \\ \rho_2 \\ \ldots \\ \rho_p \end{bmatrix} \tag{5}$$

which allow to find the coefficients $\Phi_i$ of AR(p) if ACF is known.

Another important function is so called partial autocorrelation function PAC which measures the importance of the k-th lag in the model AR (p). It is found by solving:

$$\begin{bmatrix} 1 & \rho_1 & \rho_2 \ldots & \rho_{k-1} \\ \rho_1 & 1 & \rho_1 \ldots & \rho_{k-2} \\ \ldots & \ldots & \ldots & \ldots \\ \rho_{k-1} & \rho_{k-2} & \rho_{k-3} \ldots & 1 \end{bmatrix} \begin{bmatrix} \phi_{k1} \\ \phi_{k2} \\ \ldots \\ \phi_{kk} \end{bmatrix} = \begin{bmatrix} \rho_1 \\ \rho_2 \\ \ldots \\ \rho_k \end{bmatrix} \tag{6}$$

where $\phi_{kk}$ =PAC. From Eq. 5 we see that for k=p $\phi_{ki}=\Phi_i$ and that PAC=0 for k>p. These two autocorrelation functions play an important role helping to find out whether a time series of some data is a sample of some ARIMA process and in particular of some stationary AR(p) process.

# EMPIRICAL AUTOCORRELATION FUNCTIONS.

Let us consider a sample S= {$z_1$,…, $z_n$} of some time-series. The autocorrelation function $\rho_k = \rho(k)$ can only be estimated from the data by $r_k$ :

$$r_k = \sum_{t=1}^{n-k}(z_t - \bar{z})(z_{t+k} - \bar{z}) \bigg/ \sum_{t=1}^{n}(z_t - \bar{z})^2 \qquad (7)$$

where z-bar is a standard sample mean. .

If the unknown time series is a stationary AR(p) then empirical ACF= $r_k$ should decrease quickly when k grows . To find the value of p we have to study a family of AR(k) where

$$Z_t - \varphi_{k1}Z_{t-1} - \varphi_{k2}Z_{t-2} - ... - \varphi_{kk}Z_{t-k} = a_t \qquad (8)$$

where $\varphi_{ki}$ are unknown and may be only estimated by using the Yule- Walker Eq. 6 in which $\rho_i$ are replaced by $r_i$.

$$\begin{bmatrix} 1 & r_1 & r_2... & r_{k-1} \\ r_1 & 1 & r_1... & r_{k-2} \\ ... & ... & ... & ... \\ r_{k-1} & r_{k-2} & r_{k-3}... & 1 \end{bmatrix} \begin{bmatrix} \hat{\phi}_{k1} \\ \hat{\phi}_{k2} \\ ... \\ \hat{\phi}_{kk} \end{bmatrix} = \begin{bmatrix} r_1 \\ r_2 \\ ... \\ r_k \end{bmatrix} \qquad (9)$$

The empirical PAC= $\hat{\phi}_{kk}$ and is not exactly zero for k>p but it should have a clear « cut off » at k=p. It means that it should be equal to zero within a standard error of the order of $n^{-0.5}$ where n is a sample size.

If we get a sample of some time series we cannot assume that it is a sample of some stationary AR (p). In order to discover which ARIMA model if any can fit the data we have to explore our sample with help of statistical software we have at our disposal. For a study of time series the recommended packages are S+ or R but even a popular Minitab can do. In particular we have to find:

- Histogram.
- Normal scores plot.
- Simple time series plot ($z_t$, t).
- Lagged scatter plots (($z_t$, $z_{t+k}$ ).
- Empirical ACF and PAC plots.
- Residuals after fitting plots.

If a time series is not a simple AR (p) it is not an easy task to determine exactly its structure but in many cases it can be done with high precision [22].

In the next section we will apply the empirical ACF and PAC function in order to detect a fine autocorrelation structure in some simulated data.

# A STUDY OF AN EMPIRICAL TIME SERIES

In order to prove our point that a standard data analysis makes impossible the discovery of a fine structure in time series of the data we simulated a sample of size 500 of AR (2):

$$Z_t - 0.25 Z_{t-1} - 0.5 Z_{t-2} = a_t \qquad (10)$$

where $a_t$ were normal i.i.d. with a unit variance. A standard descriptive analysis: summary, histogram, and normal scores showed that the data can be viewed as a sample drawn from some normally distributed population.

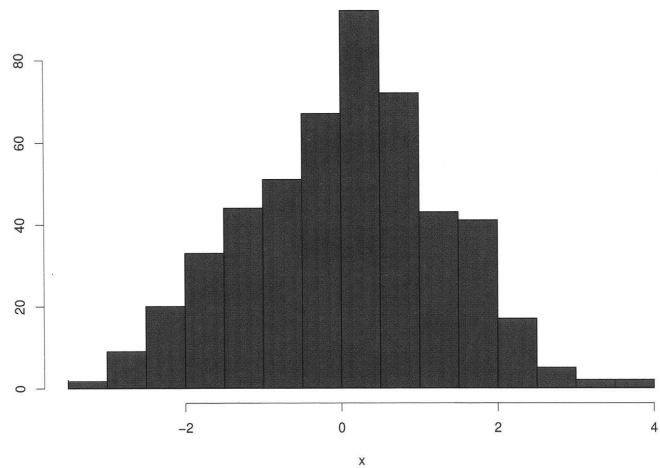

**FIGURE 1**. A Histogram

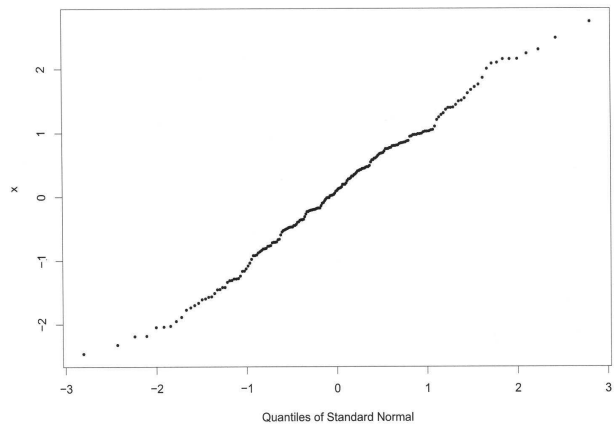

**FIGURE 2.** Normal Scores Plot

Only the detailed analysis allowed discovering the fine structure in the studied sample what can be seen from the figure below:

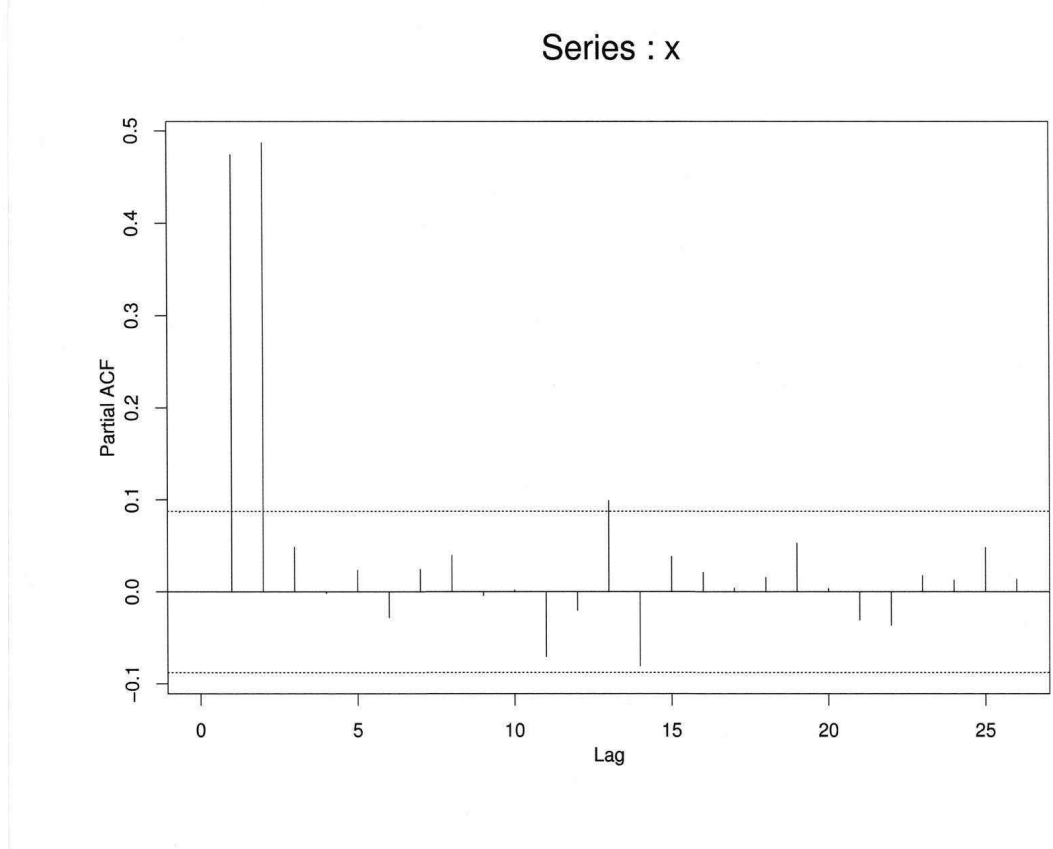

**FIGURE 3.** Empirical PAC function

The empirical ACF was decaying and the empirical PAC function had a clear <<cut off>> at lag 2 thus we could conclude that the sample was drawn from a stationary AR(2). The statistical package $S^+$ using the Eq. 6 estimated the coefficients in the Eq. 10 to be: 0.243 and 0.487 very close to the true values 0.25 and 0.5 respectively. A more detailed study of different simulated time series will be published elsewhere.

## CONCLUSIONS

Standard analysis of the data is unable to discover possible fine stochastic structures in the data. Subtle statistical analysis of the time series of the data using the tools described above and in [21, 22] is an appropriate method to test the completeness and the limitations of QT. The statements that the violation of Bell inequalities proves the non locality of QT and/or its completeness are simply not true [2,3,7,9,13,14,16,23-25] and it is really strange how often they have been repeated in the past and even during this conference.

## ACKNOWLEDGMENTS

I would like to thank Andrei Khrennikov for the warm hospitality extended to me during this interesting conference. I am indebted also to Mahmoud Zarepour from University of Ottawa who initiated me to $S^+$ and helped me with the data analysis. I would like to thank also the University of Ottawa for a financial support.

## REFERENCES

1. B.V. Gnedenko, *The Theory of Probability*, New York: Chelsea, 1962, p.40.
2. M. Kupczynski ,'' Bertrand's paradox and Bell's inequalities'', *Phys.Lett.A* **121** , 205-207(1987)
3. M. Kupczynski,''EPR paradox ,locality and completeness of quantum theory'', in *Quantum Theory Reconsideration of Foundations-4*, edited by G.Adenier et al, AIP Conf.Proc. **962**, NY:Melville 2007, pp.274-85 (arXiv:0710.3510)
4. L.E.Ballentine,'' The statistical interpretation of quantum mechanics'', *Rev.Mod.Phys*. **42**, 358(1970)
5. L..E.Ballentine, *Quantum Mechanics: A Modern Development*, Singapore:World Scientific, 1998.
6. L.Accardi, ''Topics in quantum probability'', *Phys.Rep*.**77**, 169 -92 (1981).
7. L.Accardi and M. Regoli,''Locality and Bell's inequality'',in: *QP-XIII, Foundations of Probability and Physics*, ed. A.Khrennikov, Singapore: World Scientific ,2002,pp. 1—28
8. L.Accardi and S.Uchiyama,''Universality of the EPR-chameleon model'', in *Quantum Theory Reconsideration of Foundations-4*, edited by G.Adenier et al, AIP Conf.Proc. **962**, NY:Melville, 2007, pp.15-27.
9. A.Yu. Khrennikov ,'' Bell's inequality: nonlocality, „death of reality", or incompatibility of random variables? , in *Quantum Theory Reconsideration of Foundations-4*, *edited by* G.Adenier et al, AIP Conf.Proc. **962**, NY:Melville ,2007, pp.121-127
10. A.Yu. .Khrennikov, *Contextual Approach to Quantum Formalism*, Fundamental Theories of Physics **160**, Dortrecht:Springer, 2009
11. .A.Yu .Khrennikov, *Ubiquitous Quantum Structure*, Berlin: Springer, 2010
12. M. Kupczynski, " Is Hilbert space language too rich", *Int.J.Theor.Phys*.**79**, 319(1973), reprinted in: *Physical Theory as Logico-Operational Structu*re, ed. C.A.Hooker, Dordrecht: Reidel, 1978,p.89
13. M.Kupczynski," On the completeness of quantum mechanics", arXiv:quant-ph 028061 , 2002
14. .. M. Kupczynski,'' Seventy years of the EPR paradox'' in *Albert Einstein Century International Conference* edited by J-M Alimi and A Funzfa ,AIP Conf.Proc. **861**, NY:Melville,2006, pp.516-523 (arXiv:0710.3510)
15  A.S.Holevo, *Statistical Structure of Quantum Theory*, Berlin:Springer, 2001.
16. M.Kupczynski,''Entanglement and Bell inequalities'', *J.Russ.Laser Research* **26**, 514-23(2005) (quant-phys 0407199)
17 K.Michielsen, F.Jin, H.de Raedt,''Event-based corpuscular model for quantum optics experiments'', arxiv:1006.1728[quant-phys], 2010
18. M. .Kupczynski,'' New tests of completeness of quantum mechanics', *Phys.Lett.A* **121**, p.51-53, 1987.
19. M.Kupczynski, ''On some important statistical tests'', *Riv.Nuovo Cimento* **7** , 215-27(1977)
20. .J. Gajewski and M. Kupczynski,''Purity tests for $\pi^-$d charge multiplicity distributions", *Lett.Nuovo Cimento* **26** , 81-87(1979)
21. M. Kupczynski,''Is the quatum theory predictably complete?'', *Phys.Scr*.**T135**, 014005 (2009) (arxiv :0810.1259)
22. G.E.P  Box, G.M Jenkins and G.C Reinsel, *Time Series Analysis Forecasting and Control*, Hoboken: Wiley, 2008
23. A.Yu Khrennikov and I.V Volovich,''Quantum non-locality, EPR model and Bell's theorem", in *3rd International Sakharov Conference on Physics. Proceedings*, edited. By A. Semikhatov et al., Singapore:World Scientific,2003, pp 260-267.
24. W de Baere W, "On conditional Bell inequalities…", *Lett.Nuovo Cimento* **40** , 488 (1984)
25 H. de Raedt, K.Hess and K.Mickielsen,"Extended  Boole-Bell inequalities…", arxiv:0901.2546v2